\documentclass[12pt]{iopart}

\usepackage{graphicx}
\usepackage{bm}
\usepackage{bbm} 
\usepackage{dsfont}
\usepackage{float}

\newcommand{\cut}[1]{{\hspace{-0.3em}}}

\newcommand{\cL}{\mathcal{L} }
\newcommand{\cP}{\mathcal{P} }
\newcommand{\me}{\mathrm{e}}

\begin{document}

\title[High storage capacity in the Hopfield model with auto-interactions - stability analysis]{High storage capacity in the Hopfield model with auto-interactions - stability analysis}

\author{Jacopo Rocchi}
\address{Nonlinearity and Complexity Research Group, Aston University, Birmingham B4 7ET, United Kingdom}
\ead{j.rocchi@aston.ac.uk}
\vspace{10pt}

\author{David Saad}
\address{Nonlinearity and Complexity Research Group, Aston University, Birmingham B4 7ET, United Kingdom}
\ead{d.saad@aston.ac.uk}
\vspace{10pt}

\author{Daniele Tantari}
\address{Scuola Normale Superiore, Centro Ennio de Giorgi, Piazza dei Cavalieri 3, I-56100, Pisa, Italy}
\ead{daniele.tantari@sns.it}
\vspace{10pt}

\begin{abstract}
Recent studies point to the potential storage of a large number of patterns in the celebrated Hopfield associative memory model, well beyond the limits obtained previously. We investigate the properties of new fixed points to discover that they exhibit instabilities for small perturbations and are therefore of limited value as associative memories. Moreover, a large deviations approach also shows that errors introduced to the original patterns induce additional errors and increased corruption with respect to the stored patterns.
\end{abstract}

%
%
%
%
%

\section{Introduction}

Hopfield models~\cite{refHopf1, refHopf2, refHopf3} are recurrent neural networks where connections between units form a fully connected symmetric network.
They have been proposed as models of content addressable memories, i.e. systems that are able to retrieve memory items from partial information.
Their introduction was inspired by the observation that in large physical systems, interactions between the elementary degrees of freedom generate collective phenomena, such low temperature magnetisation in Ising models. Following this idea, the stability of memories in systems of interacting neurons has been successfully described as an emergent property, instigated by the dynamics of neural network models~\cite{refHopf1, refHopf2}.

Any physical system whose dynamics is dominated by a number of locally stable states can act as a content addressable memory as long as these states can be controlled. The Hebbian rule~\cite{refHebb} has played an important role in training the couplings between neurons such that a prescribed set of memories (binary configurations of the Hopfield model) become attractors of the dynamics.
Hopfield pointed out ~\cite{refHopf1} that the issue of pattern retrieval is non trivial and that retrieval performance falls rapidly as more patterns are introduced, and are incorporated in the couplings. 
This behaviour was first found in numerical simulations and then analysed utilising statistical mechanics methods and exploiting the analogy with spin glass models.
For $P$ is the number of patterns, $N$ the number of neurons and $\alpha=P/N$, it has been found~\cite{refAmit1,refAmit2}  that the critical value $\alpha_c$ below which recovery is possible is approximately $\alpha_c  \approx 0.14$. Original studies considered the case where diagonal terms are not present, i.e. neural network models without auto-interactions. 
Subsequent studies \cite{kabashima2001tap, mezard2017mean} considered also the problem with auto-interactions, but only recently it has been pointed out \cite{folli2017maximum} that in this case an interesting regime can be found at $\alpha \gg 1$.
The probability that a given pattern is not a fixed point of the dynamics was studied and it has been shown that this probability is very small for very low $\alpha$, as expected, but surprisingly that there is another unexplored region at very large $\alpha$ where this probability is very small as well. While the former and other intermediate regimes are well studied~\cite{refAmit2}, the behaviour at $\alpha \gg 1$ was unexpected, since it implies that in this new regime the patterns are again fixed points of the dynamical equations. 
Moreover, it has been pointed out~\cite{folli2017maximum} that this new regime does not appear in the absence of diagonal interaction terms.

In this paper we study the stability of these fixed points. The relevance of this analysis comes from the fact that associative memories are useful when (in the regime where) they recover memories on the basis of similarity. In other words, the Hopfield model can be used to retrieve memories when starting the dynamics from a configuration $\underline{s}$ similar, but not exactly equal to, a given pattern, we converge to one of the original patterns. Our analysis suggests that although training patterns are fixed points of the dynamics in the newly discovered regime~\cite{folli2017maximum}, they are unstable, contrarily to the known regime of small $\alpha$ values. In Sec.~\ref{sec:Dynamics} we introduce the model and its dynamics; while in Sec.~\ref{sec:Stability} we compute the probability of escaping the stored pattern when a small perturbation is introduced. In Sec.~\ref{sec:LargeDev} we complete the analysis by computing the typical number of errors made after one dynamical step, where errors are measured in terms of the Hamming distance between the initial configuration (one of the training vectors) and the dynamical configuration.

\section {Dynamics of a neural network}
\label{sec:Dynamics}

The neural network model  that we will consider in this work is a system of $N$ binary variables (neurons) $s_i \in \{-1, +1\},~i=1 \ldots N,$ interconnected by a symmetric network of synapses specified by the real coupling matrix $J_{ij} $.
We will focus on the non-linear dynamical equations 
\begin{equation}
s_i (t +1) =  \textrm{sign} (x) \left( \sum_{j=1}^N J_{ij} s_j (t) \right)\:,
\label{eq:one}
\end{equation}
where  the value $s_i(t+1)$ represents the state of the neuron $i$ at time $t+1$, which may be active, $s_i(t+1)= +1$, or inactive, $s_i(t+1)= -1$. The value $s_i(t+1)$ depends on the state of the neurons at the previous time step, $\{s_j(t)\}$. 
These equations give rise to a dynamical process in the space of configurations, depending on the properties of the matrix $J_{ij}$ but, as pointed out by Hopfield~\cite{refHopf1}, a careful choice of $J_{ij}$ may trap the dynamics in basins of attraction that correspond to a given random set of patterns (training vectors) $\{ \underline{\xi}^{\nu} \}_{\nu=1,\ldots,P}$ where $\xi^{\nu}_{i} \in [-1, +1],~i=1 \ldots N$ and $\nu=1 \ldots P$.
These patterns can be considered as memories that the system is able to retrieve and  should be fixed points of Eq.~(\ref{eq:one}). In the following we will focus on the case where the matrix $J_{ij}$ is specified by the Hebbian rule~\cite{refHebb},
\begin{equation}
J_{ij}= \frac{1}{P} \sum_{\nu=1}^P \xi^{\nu}_i \xi^{\nu}_j\:,
\label{eq:two}
\end{equation}
introduced in order to explain \textit{associative learning}. In fact, for $P=O(N)$, Eq.~(\ref{eq:two}) can be obtained cumulatively from the successive application of the learning rule $\Delta J_{ij} \propto \xi^{\mu}_i \xi^{\mu}_j $, specifying the change in the coupling between neurons when learning a given pattern $\mu$ and describing the observation that the simultaneous activation of neurons $i$ and $j$ increased the coupling strength between them.

Retrieval of patterns is known to be possible only for a number of patterns that is a small fraction of that of the neurons~\cite{refHopf1, refHopf2,refAmit1,refAmit2}. Diagonal interaction terms were not considered in the early works about Hopfield model for a physical reason: in the corresponding spin models the field of a variable is induced by the state of its neighbours, but not on its own; thus self interactions do not exist and $J_{ii}=0,\forall i$.
Neural networks with diagonal terms have been studied in~\cite{folli2017maximum} and a very interesting regime has been found for $\alpha \gg 1$.
The probability $p_V$ that a given pattern $\underline{\xi}^{\mu}$ is a not fixed point of the dynamics has been computed and has been shown to be very small for very low $\alpha=P/N$, as expected, but surprisingly another region has been identified at the very large $\alpha$ regime, where $p_V$ is also very small. In the intermediate regime, $p_V$ is large.
The first and the intermediate regime are well known but the behaviour of $p_V$ at $\alpha \gg 1$ was new and unexpected. 
The probability $\bar{p}_V$ that a random vector, not in the training set $\{ \xi_i^{\nu} \}_{\nu=1,\ldots,P}$, is not a fixed point of the dynamics has also been studied.
As expected, in the low $\alpha$ regime, the probability $\bar{p}_V$ is close to $1$ but, and as $\alpha$ increases this probability vanishes.
Thus, in the new, large $\alpha$ regime, $\bar{p}_V$ is also very small. In other words, in this regime, both patterns $\underline{\xi}^{\nu}$ and random configurations are likely to be fixed point of the dynamics. This has a trivial interpretation by noticing that for very large $P \gg N$, the interaction matrix $J_{ij}$ defined in Eq.~(\ref{eq:two}) tends to the unit matrix.
Even if this result seems to invalidate the usefulness of this new regime, it was shown that the ratio $\bar{p}_V/ p_V$ tends to a finite number, $e$, in the large $N$ limit and that real patterns have an higher probability of being fixed points of the model. 
In the next section we address the stability question of these patterns, making use of the same strategy used in~\cite{folli2017maximum}.

\section {Stability of the fixed points}
\label{sec:Stability}

In order to study the stability of fixed points of the dynamical equations given in Eq.~(\ref{eq:one}), we  consider the case where one of the patterns is randomly perturbed. Since the patterns $\underline{\xi}^{\nu}$ are configurations of binary variables, a random perturbation is obtained by flipping the value of $K$ sites. 
We denote by $\cP$ the set of perturbed sites and by $\cL$ the set of unperturbed variables and clearly, $\cL = \overline{\cP}$.
We can consider the equations
\begin{equation}
s'_i  =   \textrm{sign} \left( \sum_{j=1}^N \sum_{\nu=1}^{P} \xi_i^{\nu} \xi_j^{\nu} s_j \right)
\label{eq:three}
\end{equation}
and compute the probability to get back to $\xi_i^{\mu}$ when the starting configuration is given by
\begin{equation}
s_j = \xi_j^{\mu} \mathds{I}_{j,\cL} - \xi_j^{\mu} \mathds{I}_{j,\cP}\:,
\label{eq:perneur}
\end{equation}
where $\mathds{I}_{j,\cP (\cL)}$ is one if $j \in \cP(\cL)$, and zero otherwise.
As in~\cite{folli2017maximum} we focus on the one-step dynamics.

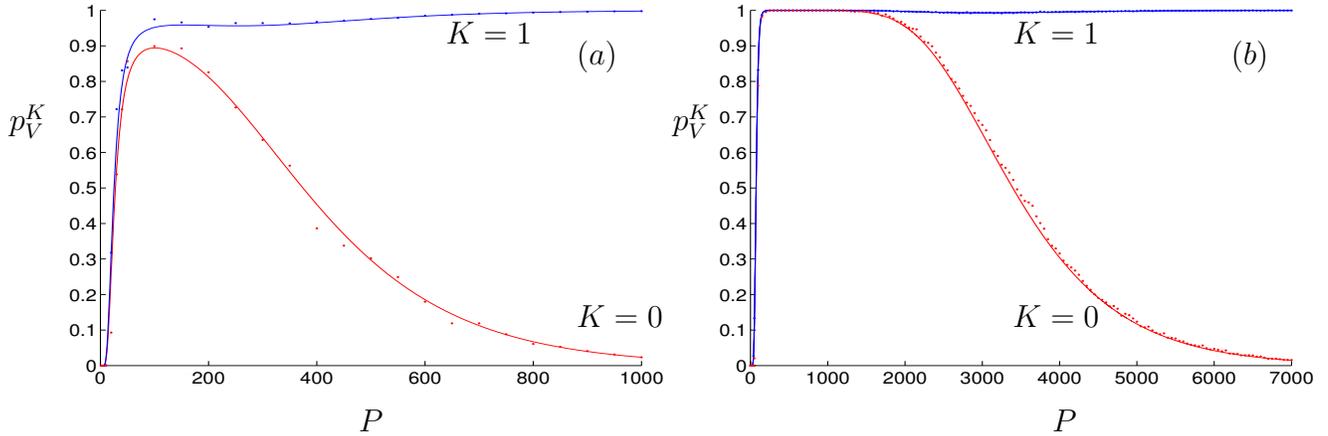
\begin{figure}[ht]
	\vspace*{0mm} \hspace*{-10mm} \setlength{\unitlength}{0.58mm}
	\begin{picture}(350,210)
	\put(149,0){\includegraphics[height=100\unitlength,width=160\unitlength]{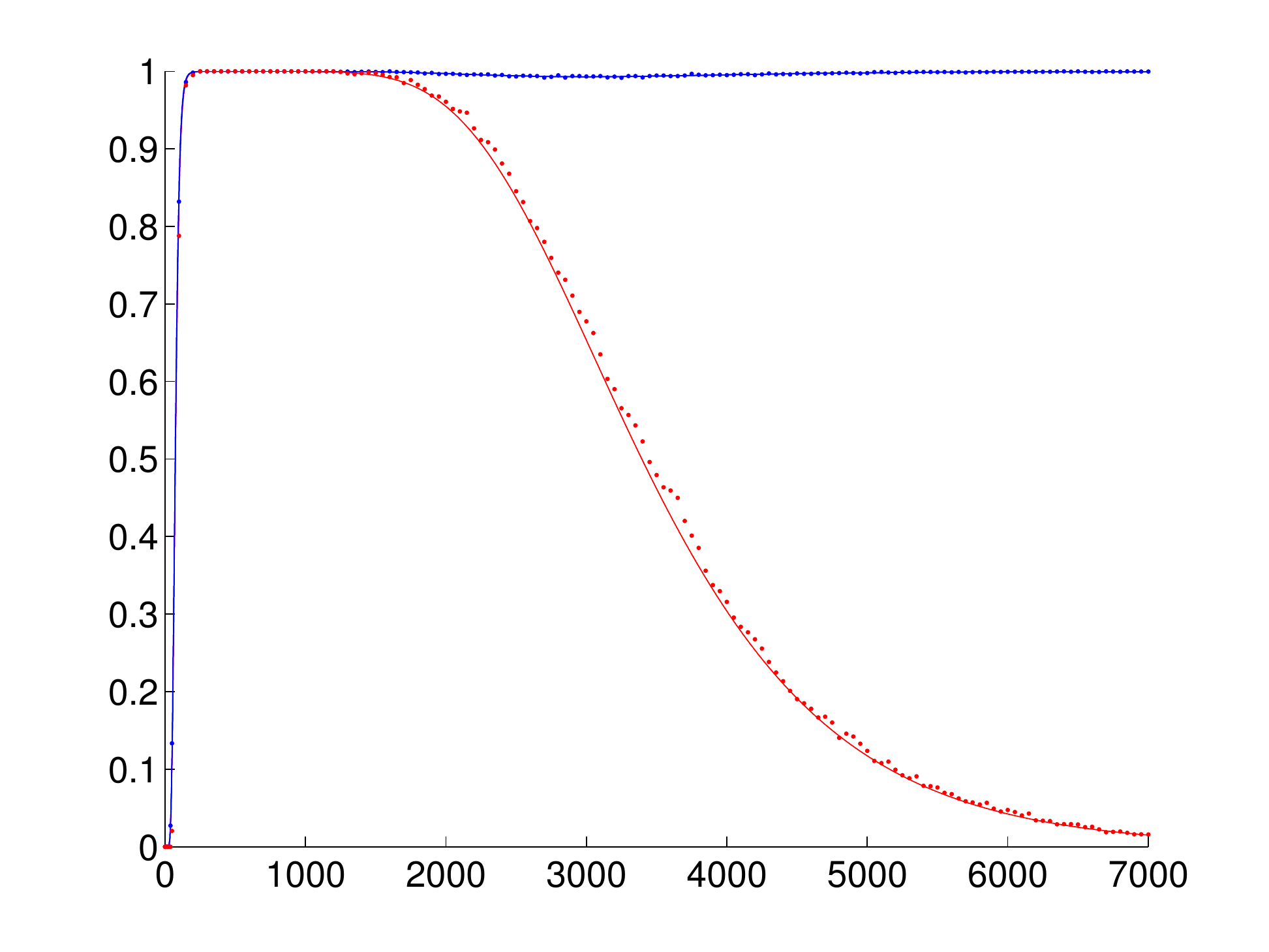}}
	\put(0,0){\includegraphics[height=100\unitlength,width=160\unitlength]{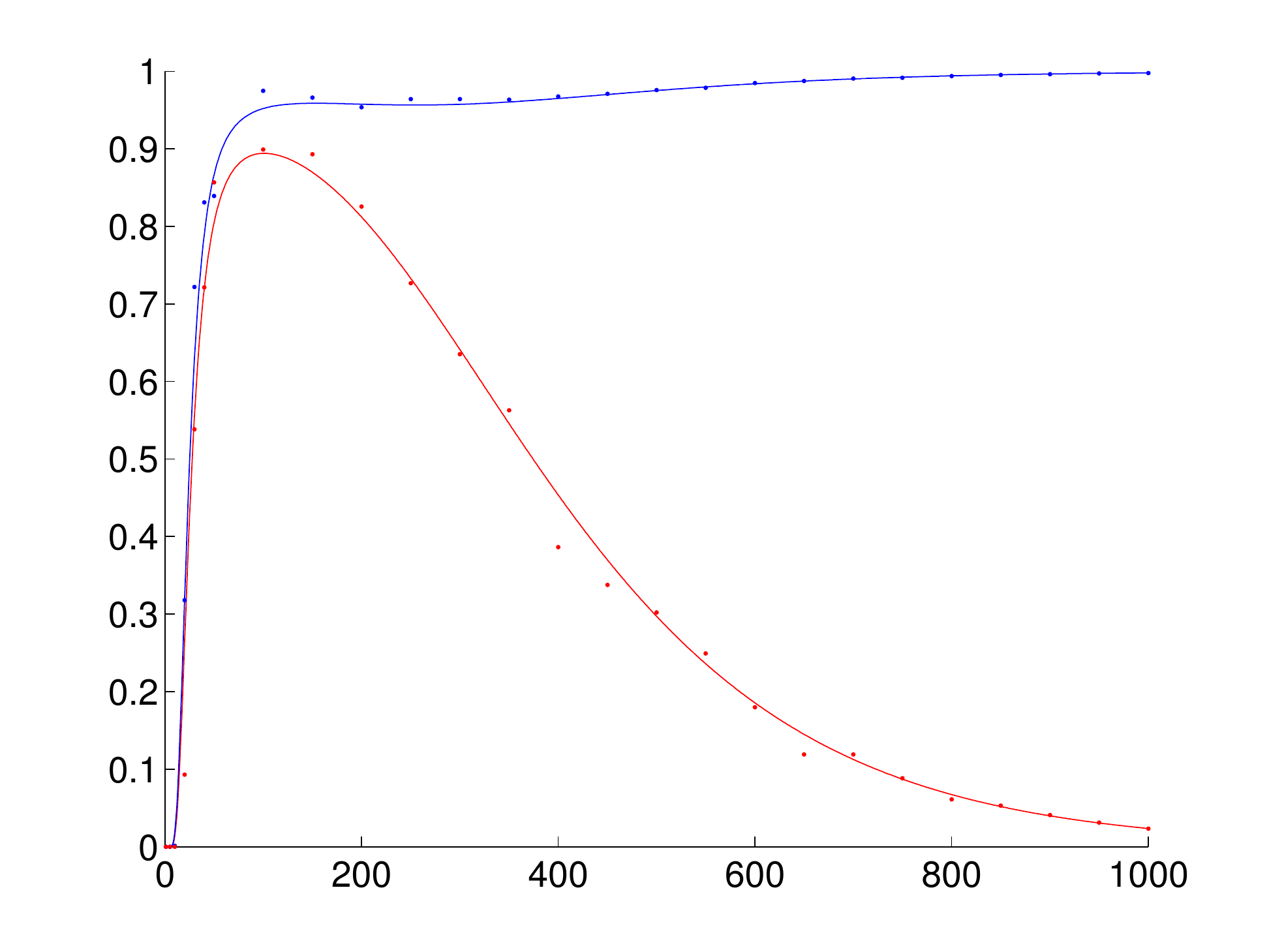}}
	\put(130,80){$(a)$}\put(280,80){$(b)$}
	\put(0,65){$p_V^{K}$}
	\put(80,-4){$P$} 
	\put(152,65){$p_V^{K}$}
	\put(239,-4){$P$} 
		\put(100,85){$K=1$} 
		\put(130,20){$K=0$} 
				\put(230,85){$K=1$} 
				\put(230,20){$K=0$} 
	\end{picture}
	\caption{The probability of not recovering the stored pattern after one iteration after $K$ variables have been perturbed $p^{K}_V $ for  $K=1$ (blue line) and $K=0$ (red line). Dots are obtained by numerical simulations, where probability $p^{K}_V $ is estimated counting the number of times that Eq.~(\ref{eq:one}) does not give a given pattern when we perturbed $K=1$ (blue dots) and $K=0$ (red dots) neurons, repeating this process $10^4$ times for different patterns at any value of $P$. (a) For $N=100$ and (b) For $N=500$.
	\label{fig_Fig1}}
\end{figure}

Let us first consider the case when $i \notin \cP$. After some elementary algebra, the argument of the $  \textrm{sign}$ function in the r.h.s. of Eq.~(\ref{eq:three}) becomes
\begin{equation}
(P+N-1-2 K ) \xi_i^{\mu} + \sum_{j\neq i} \sum_{\nu \neq \mu} \xi_i^{\nu} \xi_j^{\nu} \left [  \xi_j^{\mu} \mathds{I}_{j,\cL} - \xi_j^{\mu} \mathds{I}_{j,\cP} \right]\:.
\end{equation}
The second term contains $(N-1)(P-1)$  uncorrelated terms of unitary variance. Using the central limit theorem we obtain for large $N$  
\begin{equation}
s'_i  =   \textrm{sign} \left( \frac{(P+N-1-2 K )}{\sqrt{(N-1)(P-1)}} \xi_i^{\mu} + z \frac{}{} \right)\:,
\label{eq:onestepdyn}
\end{equation}
where $z\sim\mathcal{N}(0,1)$ is drawn from a normalised Gaussian distribution.
Clearly, if $z$ were 0, $s_i' = \xi_i^{\mu}$ and we recover the correct sign. The variable $z$, induced by the arbitrary bit flips, is thus a destabilising term, that impacts on the r.h.s. of Eq.~(\ref{eq:onestepdyn}). 
It is actually harmless as soon as it doesn't change the sign of $ \xi_i^{\mu}$, thus we make a mistake on the value $ s_i' $ with a probability equal to
\begin{equation}
p^{\cL} = \int_{-\infty}^{-\frac{(P+N-1-2K)}{\sqrt{(N-1)(P-1)}}} P(z) dz =  \frac{1}{2}   \textrm{erfc} \left \{ \frac{(P+N-1-2K)}{\sqrt{2(N-1)(P-1)}}   \right \} ~.
\label{eq:defpLdan}
\end{equation}

Analogously,  when $i \in \cP$,  the r.h.s. of Eq.~(\ref{eq:three}) becomes
\begin{equation}
(N-P+1-2K) \xi_i^{\mu} + \sum_{j\neq i} \sum_{\nu \neq \mu} \xi_i^{\nu} \xi_j^{\nu} \left [  \xi_j^{\mu} \mathds{I}_{j,\cL} - \xi_j^{\mu} \mathds{I}_{j,\cP} \right]\: .
\end{equation}
Using a central limit argument one obtains
\begin{equation}
s'_i  =   \textrm{sign} \left( \frac{(N-P+1-2K )}{\sqrt{(N-1)(P-1)}} \xi_i^{\mu} + z \frac{}{} \right)\:,
\end{equation}
where $z$ is again drawn from a normalised Gaussian distribution. 
As in Eq.~(\ref{eq:defpLdan}) we obtain the probability of making an error on one of the perturbed variables,
\begin{equation}
p^{\cP} = \int_{-\infty}^{-\frac{(N-P+1-2K )}{\sqrt{(N-1)(P-1)}}} P(z) dz =  \frac{1}{2}   \textrm{erfc} \left \{ \frac{(N-P+1-2K)}{\sqrt{2(N-1)(P-1)}}\right \}\:.
\end{equation}
Notice that  $p^{\cP}$ and $p^{\cL}$ differ in the sign in front of $P$, indicating the contribution coming from the diagonal interaction components. This contribution is always aligned with the variable value and consequently decreases the error probability in unperturbed variables.  In the limit of large $N$ and $P=\alpha N$, and finite $K$ we obtain
\begin{equation}\label{eq:limp}
p^{\cL} = \frac{1}{2}   \textrm{erfc}\left(\frac{1}{\sqrt{2\alpha}} - \sqrt{\frac{\alpha}{2}}\right)\quad\quad  p^{\cP} = \frac{1}{2}   \textrm{erfc}\left(\frac{1}{\sqrt{2\alpha}} +\sqrt{\frac{\alpha}{2}}\right).
\end{equation}
While  $p^{\cL}$  tends to zero as $\alpha \to 0$ and $\alpha\to\infty$, with a maximum at $\alpha=1$, the error probability for perturbed spins  $p^{\cP}$ is an increasing function of $\alpha$, going from $0$ to $1$. This difference in stability between perturbed and unperturbed spins affects the overall stability of the original pattern. Since there are $K$ perturbed variables and $N-K$ unperturbed variables, the probability of failing to recover the original pattern $\underline{\xi}^{\mu}$ after a single step of parallel dynamics is
\begin{equation}
p_V^{K} = 1-  \left(1-p^{\cP} \right)^{K} \left(1 - p^{\cL} \right)^{N-K}\:
\end{equation}
that, for $K=0$, becomes the probability $p_V$~\cite{folli2017maximum}.
This probability can be plotted for different $P$ values at a given $N$. In Fig.~\ref{fig_Fig1}(a) we plot $p^{K}_V $ for $N=100$ and $K=1$ (blue) and $K=0$ (red), where the second case corresponds to the unperturbed case studied in~\cite{folli2017maximum}.
We also performed numerical simulations (dots) in systems of $N=100$ variables for a different number of $P$, counting the number of times that taking one of the training vectors and perturbing $K=1$ of its values we did not recover the original training vector, repeating this procedure $10^4$ times. 
While at large $P$  values the probability $p_V$ (red line) decreases to zero, consistently with the observation made in~\cite{folli2017maximum}, this does not happen for the perturbed case (blue line). In other words, perturbing just one variable in a system of $100$ neurons is sufficient to not recover the correct patterns in the regime $P\gg N$. Moreover, we notice that for small $P$  values both lines are close to zero, meaning that in this regime perturbing one neuron does not make a big difference.  This is in agreement with
Eq.~(\ref{eq:limp}): for $K=0$,  $p_V$ follows the behaviour of $p^{\cL}$, where the stability of unperturbed spins at $K>0$ is affected by perturbed spins that are dominated by the diagonal interactions at large $P$ values, which increase the error probability.
In Fig.~\ref{fig_Fig1}(b) we plot the same quantity $p^{K}_V$ in the case $K=0$ (red line) and $K=1$ (blue line) for a system of $N=500$, finding the same qualitative behaviour.

\section{Large Deviations}
\label{sec:LargeDev}
In this section we compute the typical number of errors observed in $\underline{s}'$ with respect to the original patterns, produced by applying Eq.~(\ref{eq:three}) to the vector of dynamical variables $\underline{s}$, which is specified by Eq.~(\ref{eq:perneur}) for a given number $K$ of perturbed spins.
Let us denote by $M_\cP$ the number of errors in the set of perturbed spins $\cP$ and by $M_{\cL}$ the number of errors in the set of unperturbed spins $\cL$ in $\underline{s}$. While the probability of $M_\cP$ is given by
\begin{equation}
P_\cP(M_\cP) = {{K}\choose{M_\cP}} \left(p^{\cP}\right)^{M_\cP} \left(1-p^{\cP} \right)^{K-M_\cP}\:,
\end{equation}
the probability of $M_{\cL}$ is given by
\begin{equation}
P_{\cL}(M_{\cL}) = {{N-K}\choose{M_{\cL}}} \left(p^{\cL}\right)^{M_{\cL}} \left(1-p^{\cL} \right)^{N-K-M_{\cL}}\:.
\end{equation}
Let us denote the total number of error by $M=M_\cP + M_{\cL}$. The probability of $M$, the number of errors at the next time step, is given by
\begin{equation}
P(M) \propto \sum_{M_\cP=0}^K \sum_{M_{\cL}=0}^{N-K} P_\cP (M_\cP) P_{\cL} (M_{\cL}) \delta_{M,M_\cP + M_{\cL}} 
\end{equation}
and we readily obtain
\begin{equation}
P(M) \propto \sum_{M_\cP=\max(0,M+K-N)}^{\min(K,M)} P_\cP(M_\cP) P_{\cL}(M-M_\cP)\:.
\end{equation}
Since we are mostly interested in the large $N$ behaviour, we denote by $K = N \rho$, $M=Nm$, $M_\cP = N m_\cP$ and $M_{\cL} = N m_{\cL}$, and use the Sterling's formula for approximating the factorial of a large integer,
\begin{equation}
\lim_{N \rightarrow \infty} N ! = \sqrt{2\pi N} \left( \frac{N}{\me} \right)^N\:.
\end{equation}
Simple algebra leads to the expression
\begin{equation}
P(m) \propto \int_{\max(0,m+\rho-1)}^{\min(\rho,m)} d m_\cP \me^{N \phi (m_\cP,m)} \approx \me^{N \phi(m_\cP^{*},m)}
\end{equation}
where $m_\cP^{*}$ is the maximum of $ \phi (m_\cP,m)$ over $m_\cP \in [\max(0,m+\rho-1), \min(\rho,m)] $ at a given $m$.
The large deviation function of the probability $P(m)$ is given by $\phi(m_\cP^{*},m) \equiv \psi (m) $ and its expression is
\begin{eqnarray}
\psi(m) &=& \rho \log(\rho) + (1-\rho) \log(1-\rho) \nonumber \\
                & & + m^*_\cP \log \left( \frac{p^{\cP}}{m^*_\cP} \right) + (\rho -m^*_\cP) \log \left( \frac{1- p^{\cP}}{\rho-m_\cP^*} \right)  \nonumber \\
                & & + (m-m_\cP^*)\log \left( \frac{p^{\cL}}{m-m_\cP^*} \right) \nonumber \\
                & & + (1-\rho-m+m_\cP^*) \log \left( \frac{1 - p^{\cL} }{1-\rho-m+m_\cP^*} \right) \:.
\end{eqnarray}
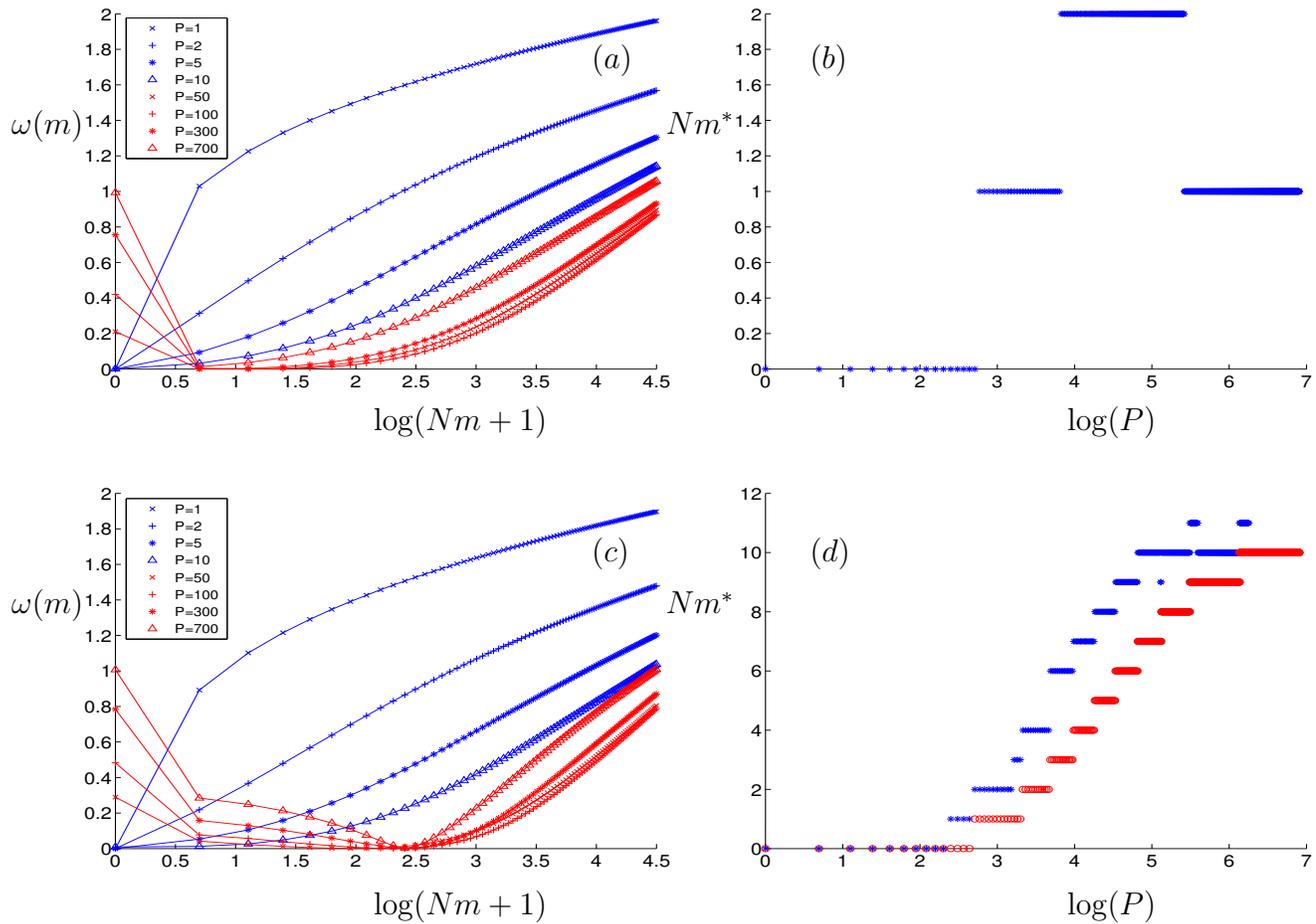
\begin{figure}[ht]
	\vspace*{0mm} \hspace*{-10mm} \setlength{\unitlength}{0.58mm}
	\begin{picture}(350,210)
	\put(149,110){\includegraphics[height=100\unitlength,width=160\unitlength]{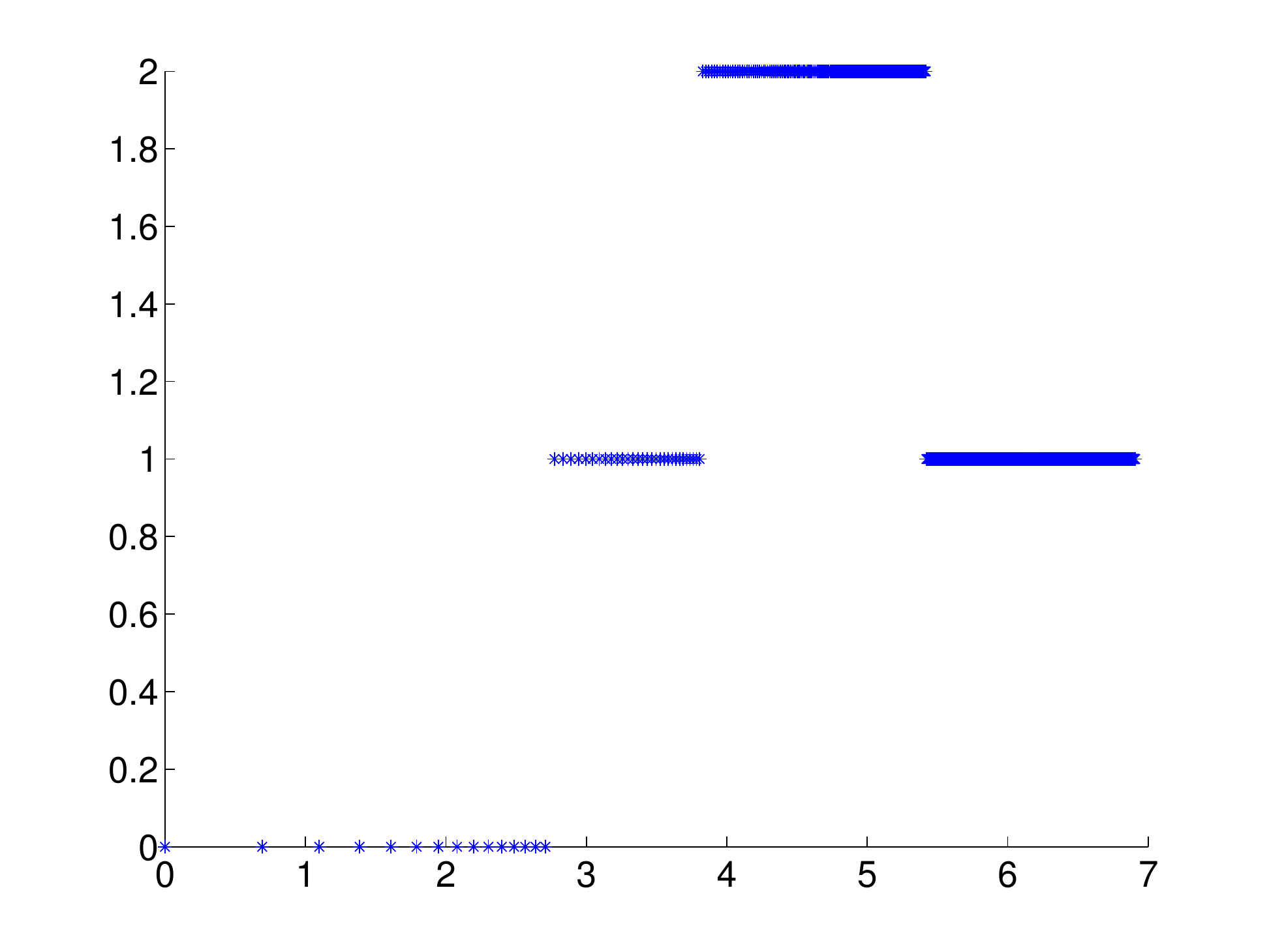}}
	\put(0,110){\includegraphics[height=100\unitlength,width=160\unitlength]{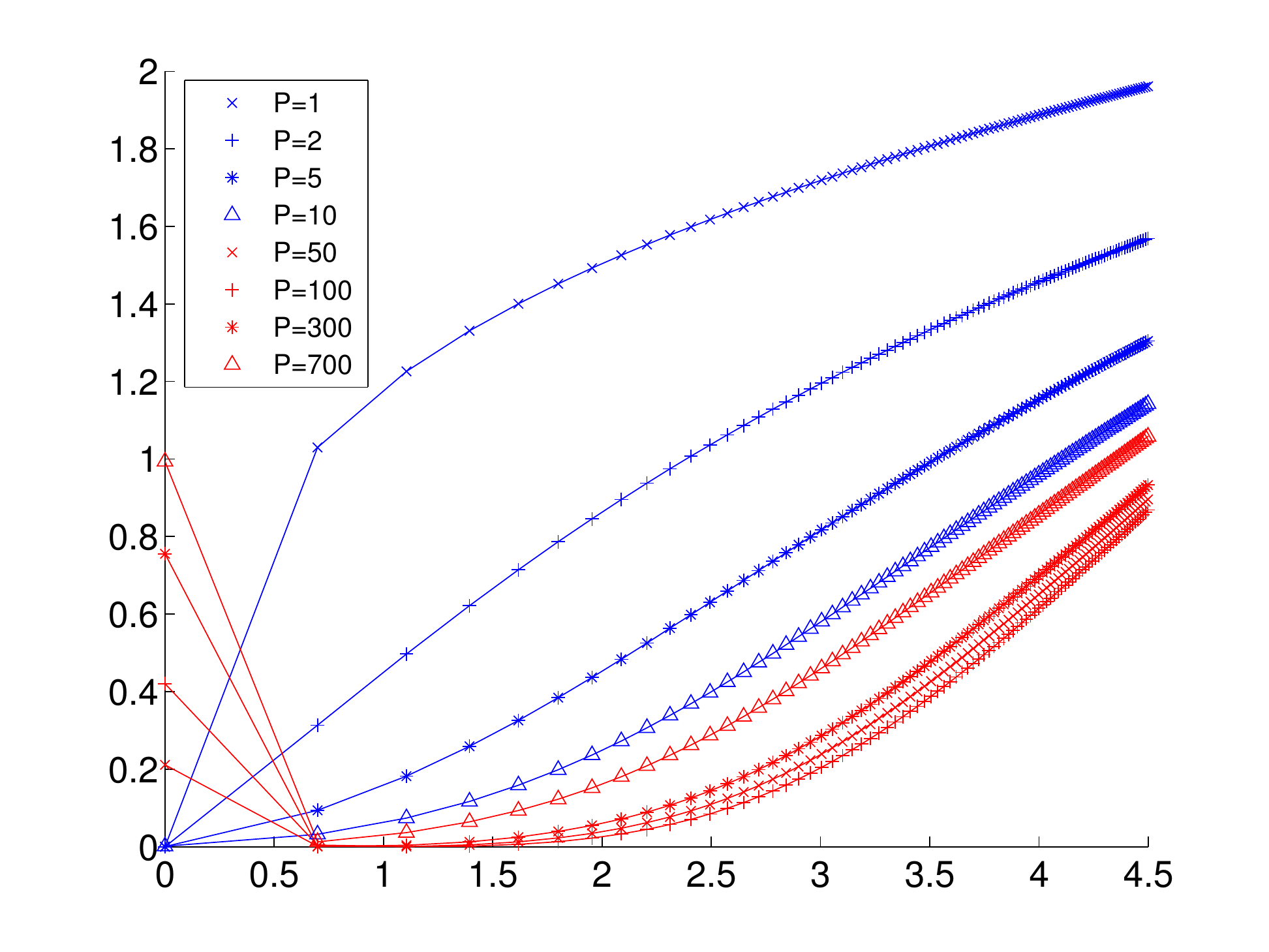}}
	\put(149,0){\includegraphics[height=100\unitlength,width=160\unitlength]{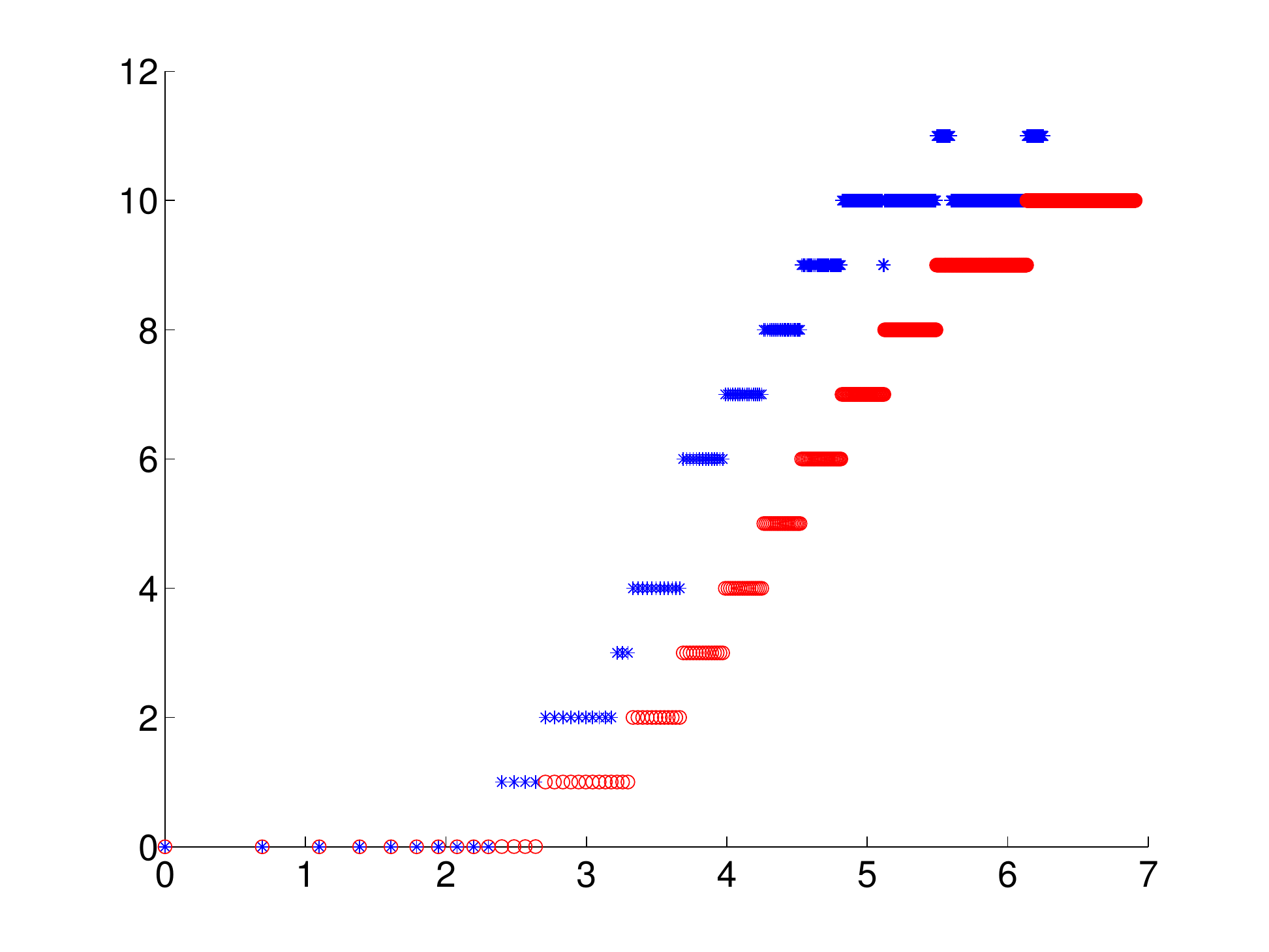}}
	\put(0,0){\includegraphics[height=100\unitlength,width=160\unitlength]{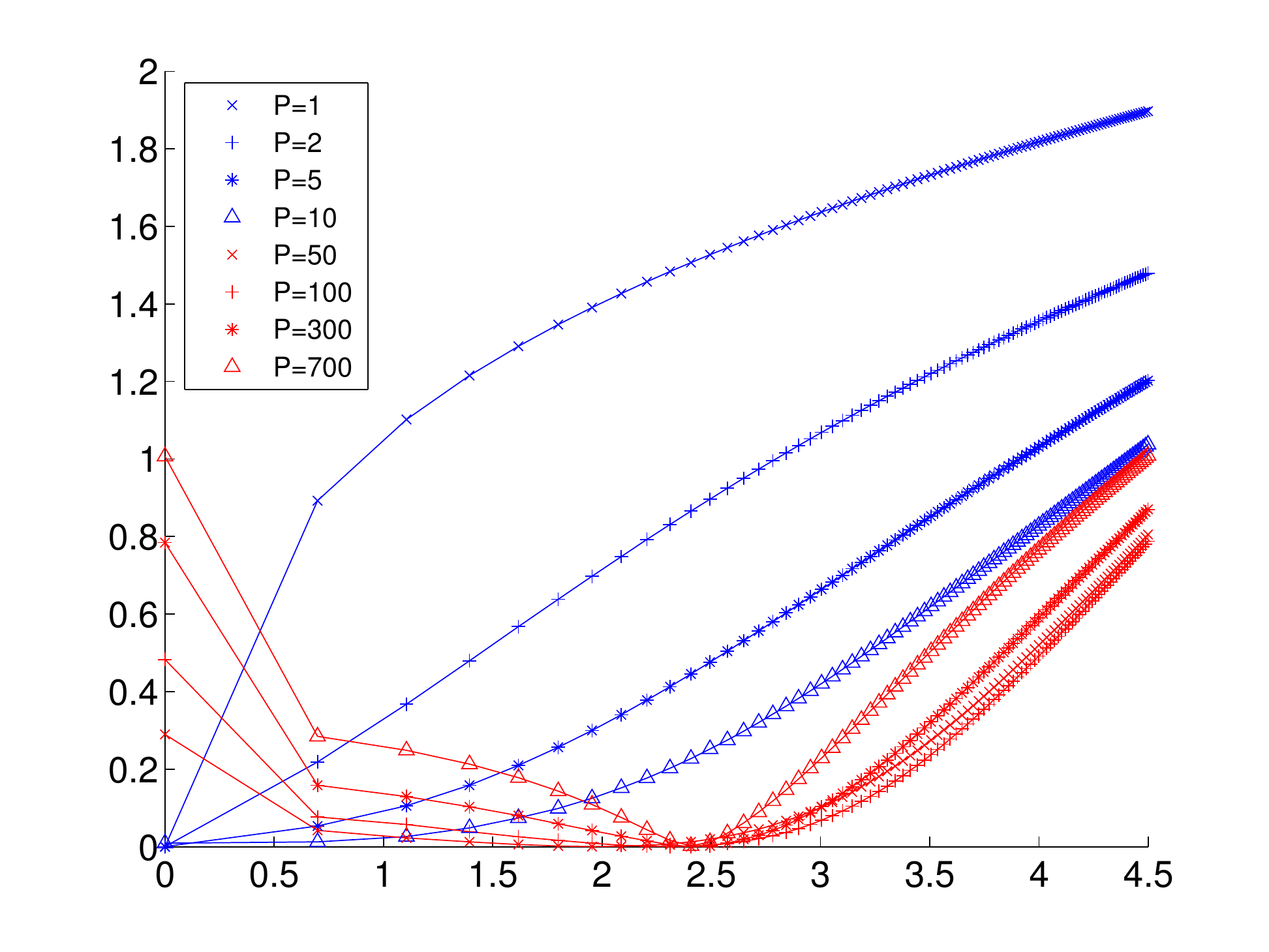}}
	\put(130,190){$(a)$}\put(180,190){$(b)$}\put(130,77){$(c)$}\put(180,77){$(d)$}
	\put(-3,65){$\omega(m)$}
	\put(-3,175){$\omega(m)$}
	\put(80,-4){$\log (N m+1)$} 
	\put(80,107){$\log (N m+1)$}
	\put(147,175){$N m^{*}$}
	\put(147,65){$N m^{*}$}
	\put(239,-4){$\log( P)$} 
	\put(239,107){$\log( P)$}
	\end{picture}
\caption{(a) Function $\omega(m)$, defined in Eq. ~(\ref{eq:omega}), for different values of $P$ at $N=100$, $K=0$, plotted as a function of $\log(N m+1)$. The minimum of this function corresponds to the most likely values of $N m$, i.e. the typical number of errors made after one dynamical step. (b) The dominating parameter $N m^*$ for $N=100$, $K=0$ for different (logarithmic) values of $P$. We see that in the large $P$ regime $ N m^{*}>0$ even when no perturbations are introduced. In this regime, the typical number of mistakes made is $N m^{*} = 1$. Notice the severe flattening of the large deviation function observed in sub-figure (a) for values $Nm > 0$. (c) The same as (a) but for $K=10$. (d) The blue x-marks indicate the dominating parameter plotted in (b) $N m^*$  for $K=10$, while the red circles indicate $N m^*_{\cP}(m^*)$, i.e. the typical number of error made in the perturbed set of variables. For small values of $P$ the dominating value is $m^*=0$, implying that we correctly reduce the number of errors just after one dynamical time step and recover the original pattern with probability $1$. However, in the large $P$ regime, $ N m^* = 10 $ implying that the number of errors is not reduced at all. }
\label{fig_Fig2}
\end{figure}
Let us first consider the case $\rho=0$, the case where the selected pattern is not perturbed at all. In Fig.~\ref{fig_Fig2}(a) we plot 
\begin{equation}
\omega(m) = \log(\log(-\psi(m)+1)+1)
\label{eq:omega}
\end{equation}
for $N=100$, $K=0$ as a function of $\log(m+1)$ for different choices of $P$.
The maximum $m^*$ of $\psi(m)$ corresponds to a minimum of $\omega(m)$, where the double logarithm is chosen to emphasise the difference between different lines at large $P$ values. 
While at small $P$ values we find $m^*=0$, i.e. the most likely value for $m$ is zero, corresponding to a non-increasing number of errors, as $P$ increases the probability of observing $m=0$ decreases and for $P>15$ (which corresponds roughly to $\alpha=0.15$) we find a different minimum at $m>0$, as can be seen in Fig.~\ref{fig_Fig2}(b) and Fig.~\ref{fig_Fig3}(a).
We also observe that  the probability of $N m$ is sharply peaked in $0$ in the low $P$ regime, while it is much broader in the large $P$ regime, even if it is clearly visible that the probability of observing $N m=0$ is negligible.

The behaviour of $m^*$ at $K=10$ can be seen in Fig.~\ref{fig_Fig2}(d) leading to a qualitatively similar behaviour of $\omega(m)$ shown in Fig.~\ref{fig_Fig2}(c), where we observe that small values of $P$ are dominated by $m^*=0$, while as $P$ increases $m^*$ remains grater than zero. In other words, while the low $P$ regime leads to a recovery of the original pattern with probability $1$ even in the case when we perturb $K=10$ variables, the large $P$ regime does not. 
Notice in Fig.~(\ref{fig_Fig2})(d) that the value of $m^*$ is mainly given by $m_{\cP}^*$ with fluctuations of order $1/N$: errors are mainly induced by the set of perturbed spins that remain blocked because of the dominating diagonal terms. 

Finally, to emphasise the sensitivity of fixed points to perturbations we plotted the $m^*$ values for a single pattern error $K=1$ shown in Fig.~\ref{fig_Fig3}(b). We observe the same qualitative behaviour of Fig. ~\ref{fig_Fig2}(b). 

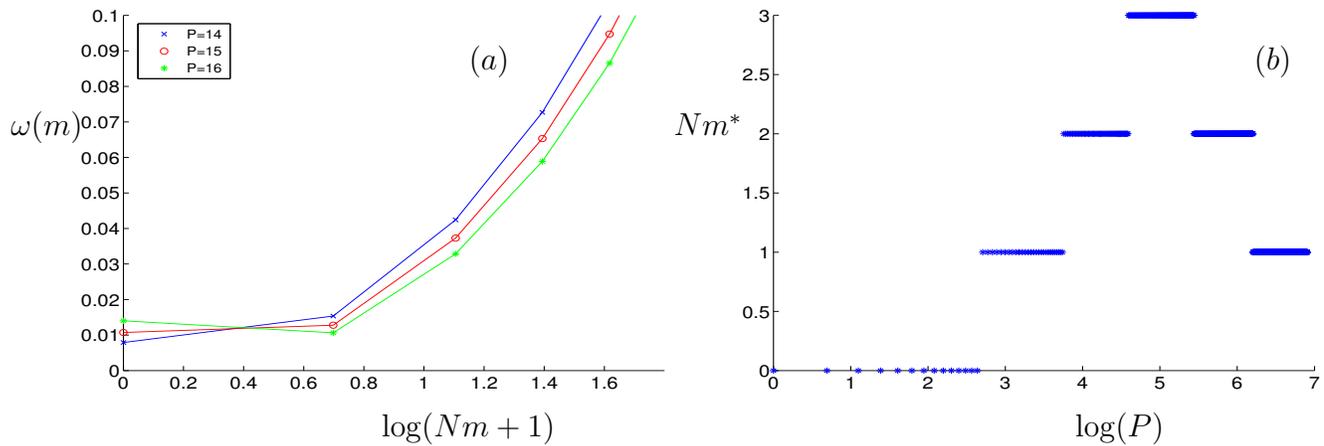
\begin{figure}[ht]
	\vspace*{0mm} \hspace*{-10mm} \setlength{\unitlength}{0.58mm}
	\begin{picture}(350,210)
	\put(149,0){\includegraphics[height=100\unitlength,width=160\unitlength]{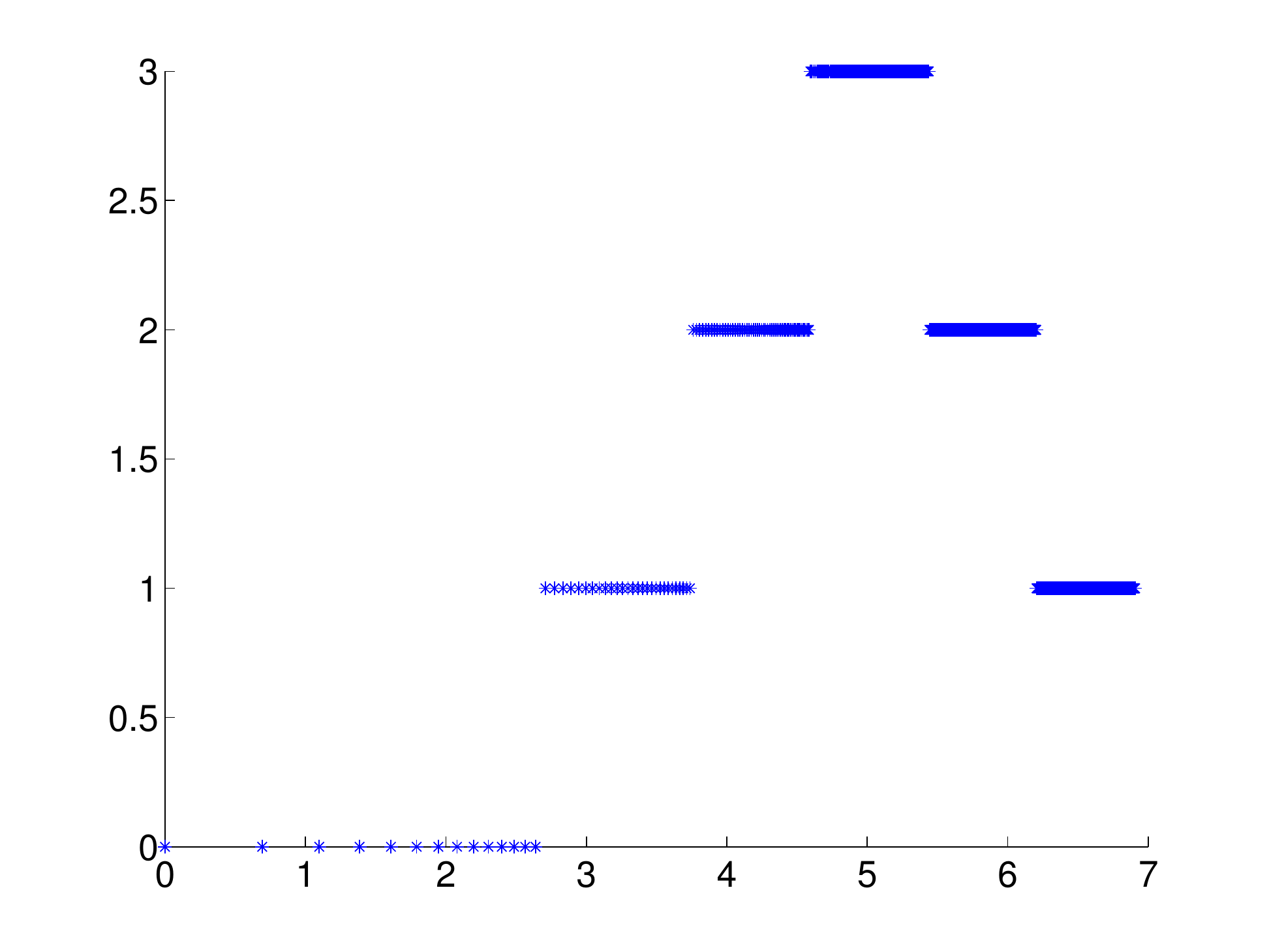}}
	\put(0,0){\includegraphics[height=100\unitlength,width=160\unitlength]{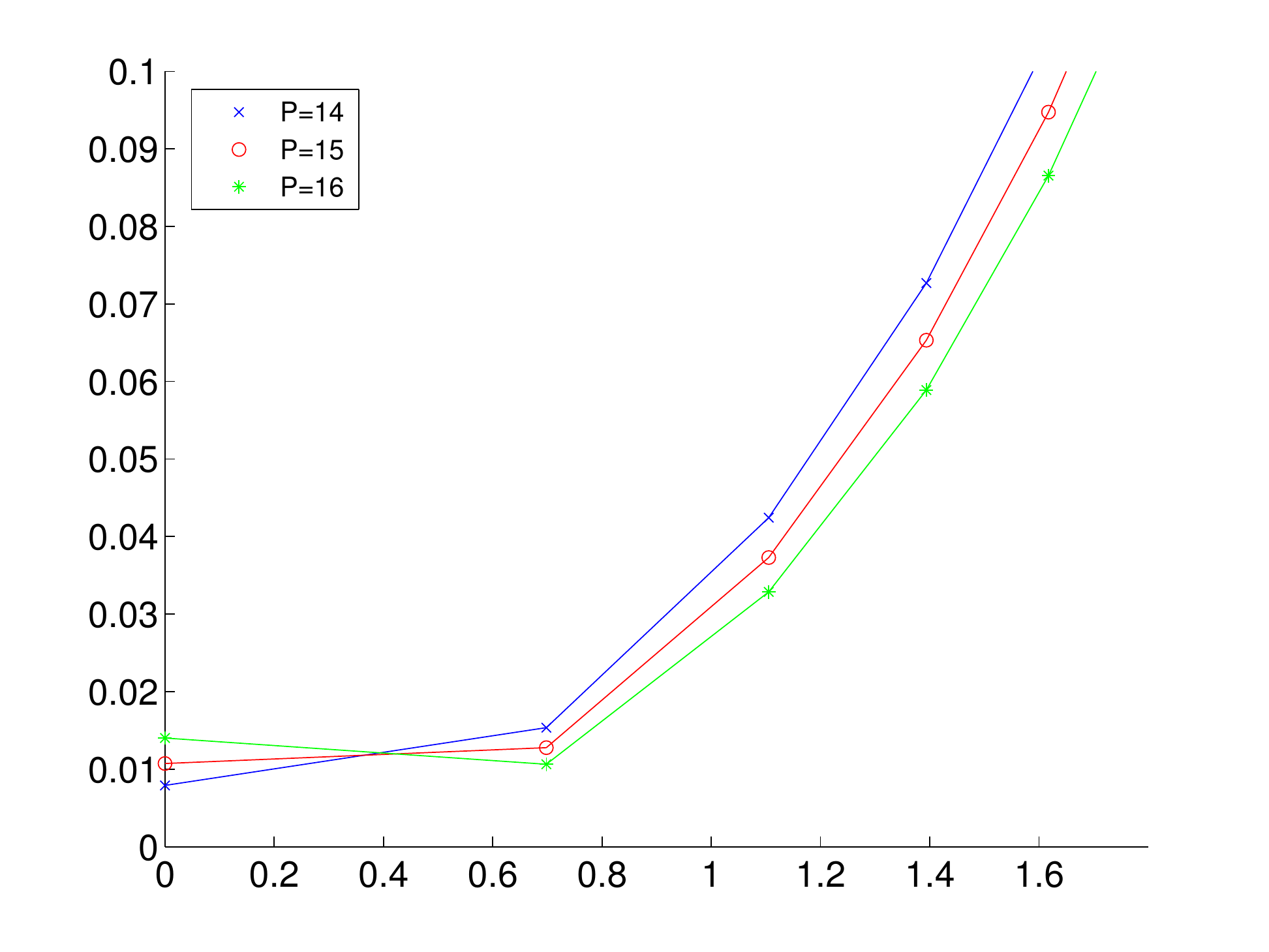}}
    \put(100,80){$(a)$}\put(280,80){$(b)$}
	\put(-5,65){$\omega(m)$}
	\put(80,-4){$\log (N m +1)$} 
	\put(147,65){$N m^{*}$}
	\put(239,-4){$\log (P)$} 
	\end{picture}
	\caption{(a) Function $\omega(m)$, defined in Eq. (\ref{eq:omega}), for different values of  $P=14,15,16$ at $N=100$, $K=0$, plotted as a function of $\log(N m+1)$. We see that for $P>15$ the minimum of $\omega(m)$ becomes non-zero. (b) The dominating parameter $N m^*$ for $N=100$, $K=1$ for different values of $P$. The $x$-axis is $\log(P)$. We basically gain the same information we got from Fig. \ref{fig_Fig2}(b): in the large $P$ regime the typical number of errors is greater than zero.}
	\label{fig_Fig3}
\end{figure}

\section*{Conclusion}

The discovery of fixed points in the Hopfield model at the large number of patterns limit raised new hopes for a high-capacity properties of the Hopfield model, especially in within the context of associative memories in neural networks. We examine the usefulness of the newly discovered fixed points by focussing on their ability to recover stored patterns on the basis of incomplete information. In other words, the ability to converge to the original pattern when starting from a configuration that is similar to, but not exactly equal to it.
We study the stability properties of these fixed points and show that these fixed points are unstable with respect to small perturbations. We also investigate the typical number of errors made by the one time step dynamics given in Eq. (\ref{eq:one}) and find that while this number is zero in the low storage regimes, it is not in the new large storage regime.

Finally, we notice that a simple statistical mechanical argument suggests that it is unlikely that the phase diagrams of an Hopfield model with auto-interactions differs from the phase diagram of an Hopfield model without auto-interactions.
In fact, the partition functions of an Hopfield model with auto-interactions and that of an Hopfield model without auto-interactions differ from sub-leading terms in $N$ and so their thermodynamical properties have to be the same.
Thus the presence of a new, unexplored, thermodynamical phase comprising multiple stable fixed points with non-vanishing basins of attraction at $\alpha \gg 1$ has to be ruled out. 

\section*{Acknowledgement}
Support from The Leverhulme Trust grant RPG-2013-48 is acknowledged. We wish to thank Pierfrancesco Urbani for insightful discussions.

\section*{Bibliography}

\providecommand{\noopsort}[1]{}\providecommand{\singleletter}[1]{#1}%


\begin{thebibliography}{1}

\bibitem{refHopf1}
J~J Hopfield.
\newblock Neural networks and physical systems with emergent collective
  computational abilities.
\newblock {\em Proceedings of the national academy of sciences},
  79(8):2554--2558, 1982.

\bibitem{refHopf2}
J~J Hopfield.
\newblock Neurons with graded response have collective computational properties
  like those of two-state neurons.
\newblock {\em Proceedings of the national academy of sciences},
  81(10):3088--3092, 1984.

\bibitem{refHopf3}
J~J Hopfield, D~I Feinstein, and R~G Palmer.
\newblock {'Unlearning'} has a stabilizing effect in collective memories.
\newblock {\em Nature}, 304:158 -- 159, 1983.

\bibitem{refHebb}
D~O Hebb.
\newblock {\em The organization of behavior: A neuropsychological theory}.
\newblock Psychology Press, 2005.

\bibitem{refAmit1}
D~J Amit, H~Gutfreund, and H~Sompolinsky.
\newblock Spin-glass models of neural networks.
\newblock {\em Physical Review A}, 32(2):1007, 1985.

\bibitem{refAmit2}
D~J Amit, H~Gutfreund, and H~Sompolinsky.
\newblock Storing infinite numbers of patterns in a spin-glass model of neural
  networks.
\newblock {\em Physical Review Letters}, 55(14):1530, 1985.

\bibitem{kabashima2001tap}
Y~Kabashima and D~Saad.
\newblock The {TAP} approach to intensive and extensive connectivity systems.
\newblock {\em Advanced Mean Field Methods--Theory and Practice}, 6:65--84,
  2001.

\bibitem{mezard2017mean}
M~M{\'e}zard.
\newblock Mean-field message-passing equations in the hopfield model and its
  generalizations.
\newblock {\em Physical Review E}, 95(2):022117, 2017.

\bibitem{folli2017maximum}
V~Folli, M~Leonetti, and G~Ruocco.
\newblock On the maximum storage capacity of the hopfield model.
\newblock {\em Frontiers in Computational Neuroscience}, 10:144, 2017.



\end{thebibliography}
\end{document}